\documentclass[aps,pre]{revtex4}

\usepackage{graphicx}

\usepackage{enumerate}
\usepackage[monochrome]{color}

\usepackage{amssymb,amsfonts,amsmath,color}

\begin{document}

\title{Fixation and absorption in a fluctuating environment}

\author{Matan Danino}
\affiliation{Department of Physics, Bar-Ilan University,
Ramat-Gan IL52900, Israel}

\author{Nadav M. Shnerb}
\affiliation{Department of Physics, Bar-Ilan University,
Ramat-Gan IL52900, Israel}

%%%%%%%%%%%%%%%%%%%%%%%%%%%%%%%%%%%%%%%%%%%%%%%%%%%%%%%%%%%%%%%%

\begin{abstract}
\noindent   A fundamental problem in the fields of population genetics, evolution, and community ecology, is the fate of a single mutant, or invader, introduced in a finite population of wild types. For a fixed-size community of $N$ individuals, with Markovian, zero-sum dynamics driven by stochastic birth-death events, the mutant population eventually reaches either fixation or extinction.  The classical analysis, provided by Kimura and his coworkers, is focused on the neutral case,  [where the dynamics is only due to  demographic stochasticity (drift)], and on  \emph{time-independent} selective forces (deleterious/beneficial mutation). However,  both theoretical arguments and  empirical analyses  suggest that in many cases the selective forces fluctuate in time (temporal environmental stochasticity). Here we consider a generic model for a system with demographic noise and fluctuating selection. Our system is characterized by the time-averaged (log)-fitness $s_0$ and zero-mean fitness fluctuations. These fluctuations, in turn, are parameterized by their amplitude $\gamma$ and their correlation time $\delta$. We provide asymptotic (large $N$) formulas for the chance of fixation, the mean time to fixation and the mean time to absorption. Our expressions interpolate correctly between the constant selection limit $\gamma \to 0$  and the time-averaged neutral case $s_0=0$.
\end{abstract}

\maketitle

\section{introduction}

Complex systems are usually affected by both deterministic and stochastic forces, and a reliable assessment of their relative importance is, in many cases, a difficult task. The neutralist-selectionist debate \cite{nei2005selectionism} in the field of molecular biology is a typical example: selectionists believe that deterministic selection is the dominant mechanism that shapes the genetic polymorphism in a population, while neutralists stress the effect of demographic stochasticity (drift). The neutral model (with some modifications, like spatial structure) was imported to ecology by Hubbell \cite{Hubbell2001unifiedNeutral,leigh2007neutral}, and the arguments about the relative importance of deterministic (niche) vs. stochastic (neutral) factors have filled many pages of the ecological literature ever since \cite{mcgill2006empirical,TREE2011,ricklefs2012global}.

 In these debates, the effect of deterministic forces is usually contrasted with \emph{demographic stochasticity}, that is, those random aspects of dynamics that affect the reproductive success of individuals  in an \emph{uncorrelated} (between individuals and over time) manner. On the other hand, the  selective/niche forces are assumed to affect an entire population (species, allele, phenotype, strain) and to be independent of time.

Recently, many authors have  considered  another possibility:  fluctuating selective pressure, or temporal environmental stochasticity \cite{lande2003stochastic,ashcroft2014fixation,cvijovic2015fate,kalyuzhny2015neutral,hidalgo2017species}. Time-varying environment may affect the selective advantage of an entire population, adding to the model a force which is correlated among individuals of the same type but changes randomly through time.

There are several good reasons to engage in models that allow for temporal environmental stochasticity. A-priori, it is difficult to imagine a mutation or a trait which are purely beneficial. An increase of body mass, for example,  may have many beneficial aspects but  it exposes the individual to an increased pressure when the environment deteriorates (e.g., during a drought). These trade-offs are quite ubiquitous  in nature~\cite{cvijovic2015fate} so one expects  environmental variations to change the relative fitness of  species and strains. Moreover, the per-generation variations in population size due to environmental stochasticity  are ${\cal O}(n)$ (where $n$ is the size of the population), while demographic stochasticity generates ${\cal O}(\sqrt{n})$ noise. Therefore, for a population of a reasonable size environmental stochasticity should be the dominant process \cite{lande2003stochastic}.

Empirically, the fluctuations in population size that were measured  in a wide variety of systems  scale in many cases like $n$, and in almost any case were found to be much larger than $\sqrt{n}$~\cite{leigh2007neutral,kalyuzhny2014niche,kalyuzhny2014temporal,chisholm2014temporal}. Measurements of the selection coefficients for different characters (or variants) of a single species also indicate that selective forces are time dependent, often changing their direction \cite{bell2010fluctuating}. Consequently, the need to extend the theory in order  to incorporate environmental stochasticity (also known as temporal niches, fluctuating selection, alternating selective pressure and so on) received a considerable  attention during  the last years \cite{kessler2014neutral,kessler2015neutral,saether2015concept,cvijovic2015fate,
kalyuzhny2015neutral,danino2016effect,danino2016stability,
fung2016reproducing,hidalgo2017species,danino2017environmental}.

In this paper we would like to address a simple question, which is also one of the cornerstones of the theories of population genetics and community dynamics: the fate of a single mutant in a finite size community. This question has been addressed long ago for the cases with pure demographic stochasticity  and constant selection \cite{crow1970introduction,ewens2012mathematical}, and we would like to extend the theory to include random selective forces. To do that,  we consider a simple and generic model for a community of $N$ individuals,  affected by selection, demographic noise and temporal environmental stochasticity. Using asymptotic (large $N$) techniques we obtained  expressions for three quantities:

\begin{enumerate}
  \item The chance of fixation for a single mutant, $\Pi(n=1)$, which is the probability that the system ends up in the absorbing state with $N$ mutants.

  \item The average time to absorption (fixation or loss) $T_A(n=1)$,  the expected time taken to reach any one of the absorbing states, given that the system is started with a single mutant/invader.

  \item The average fixation time $T_f(n=1)$, i.e. the mean time between the introduction of the mutant/invader and the fixation of the system by its lineage, \emph{conditioned on fixation}.

\end{enumerate}

In the above definitions, average is taken   over both histories \emph{and}  initial conditions. For example, the average time to absorption is defined as $T_A = (T_A^+ + T_A^-)/2$, where $T_A^+$   ($T_A^-$)  are the average time to absorption when at $t=0$ the environment was in the plus (minus) state, i.e., when the fitness of the mutant type is higher (lower) than that of the wild type (see formal definitions below).

\section{Methods}

We consider a community of $N$ individuals, where at $t=0$ one individual is a mutant or invader and all others are wild types. Our model is inspired by the standard competitive Lotka-Volterra dynamics, where two species compete for the same resource. In its individual based version one may consider two (randomly picked) individuals that fight for a piece of food, say, the winner reproduces and the loser dies. In each elementary step of our Moran process two individuals ($i$ and $j$) are chosen at random for such a duel. If both individuals belong to the same species, the result of the duel does not affect the abundance. In case of an interspecific duel, the chance of an individual to win depends on its relative fitness.   The mutant and its descendants have logarithmic fitness $s_\mu$ and  wild type individuals have fitness  $s_w$. A mutant type wins a duel against a wild type with probability,
 \begin{equation}
      P_{\mu} = \frac{1}{2} + \frac{s_\mu - s_w}{4},
\end{equation}
where the chance of the wild type to win is $1-P_{\mu}$. Since $P_{\mu}$ depends only on $s_\mu - s_w$, we can take, without loss of generality, $s_w=0$  and denote $s_\mu$ (the logarithmic relative fitness of the mutant) simply by $s$. Time is measured in units of generations, where a generation is defined as $N$ elementary duels.

Under temporal environmental stochasticity $s$ is a function of time and we assume that it takes the form
\begin{equation}
s(t) = s_0+\eta(t),
 \end{equation}
 where $s_0$ is the time-averaged (log)-fitness difference between the mutant lineage and the wild types while the (zero mean) variable $\eta(t)$ reflects the effect of environmental variations. These environmental fluctuations are characterized by two quantities: their amplitude $\gamma$ and their correlation time (measured in units of a generation) $\delta$.

  Following \cite{danino2016stability, hidalgo2017species}  we model temporal environmental stochasticity by dichotomous (telegraphic) noise, so $\eta(t)$ may take two values, either $(+\gamma)$ or $(-\gamma)$. After each elementary duel the chance of the environment to stay in the same state is $1-1/(\delta N)$, while its chance to flip (i.e, $\pm \gamma \to \mp \gamma$) is $1/(\delta N)$. Both white Gaussian noise and white Poisson noise can be recovered from the dichotomous noise by taking suitable limits~\cite{ridolfi2011noise}, so the results obtained here are quite generic. A detailed description of the process, including the transition probabilities and the form of the corresponding backward Kolomogorov equation (BKE), is given in Appendix \ref{apa}.

  Up to this point, our analysis is very similar to the one presented for the same problem in \cite{ashcroft2014fixation}, who provided closed-form expressions for  fixation times and the fixation probability using the theory of Markov chains and the elementary transition rates. Here we would like to obtain explicit and simple expressions for these quantities in the  large-$N$ limit, which is the relevant regime in most of the realistic applications.

  To do that, we implement the techniques we have developed recently in \cite{danino2016stability}  (the main results that are relevant to this work are summarized in Appendix \ref{apa}). We used the continuum approximation, where  the number of mutants $n$ is replaced  by their fraction $x = n/N$ and quantities like $\Pi(x+1/N)$ are expanded to second order in $1/N$.  The relevant BKEs emerge as two coupled, second order differential equations [such as Eqs. (\ref{eq61a}) below]. Using a  dominant balance analysis we can show that, in the large $N$ limit, these BKEs may be reduced to a \emph{single} second order differential equation. This procedure is demonstrated in Appendix \ref{apa} for the time to absorption $T_A \equiv (T_A^+ + T_A^-)/2$: instead of having two coupled equations for $T_A^{\pm}$, we obtain a single equation for $T_A$.

 Using that, and the standard techniques to obtain $\Pi$, $T_A$ and $T_f$ \cite{redner2001guide}, we can write down, for each case, the relevant equation with the appropriate boundary conditions, as detailed in the appendices below [Eq. (\ref{eq63}), Eq. (\ref{eq90}) and the pair of equations (\ref{eq200a}) and (\ref{eq200})]. In all three cases the set of equations may be solved quite easily using integration factor, but the results are given in terms of nested  integrals over hypergeometric functions that do not provide a transparent analytic picture.  To overcome this difficulty, we have calculated the leading terms in the large $N$ asymptotic series.

 The details of these calculations are given in the three appendices \ref{apb},\ref{apc} and \ref{apd}.  In the  next section we present and discuss the bottom-lines results  in terms of $s_0$ (the time-averaged mutant fitness), $N$ (that sets the scale of demographic noise, which is $1/N$), $g = \gamma^2 \delta/2$, the effective strength of environmental stochasticity, $\alpha = s_0/g$, the ratio between deterministic and stochastic selective forces  and $G \equiv Ng$, the ratio between the environmental and the demographic stochasticity (see glossary).
 \begin{table}
\begin{center}
\caption {Glossary} \label{table1}
    \begin{tabular}{ | l |  p{10cm} |}

    \hline
    Term  &  Description \\ \hline
    $N$ &  number of individuals in the community. \\ \hline
    $n$ &  number of mutant type individuals ($N-n$ wild type). \\ \hline
    $x$ &  fraction of mutants, $x=n/N$.  ($1-x$ is the fraction of wild type)\\  \hline
    $\delta$ & correlation time of the environment, measured in  generations.\\
    \hline
    $s_0$ & time-averaged  fitness of the mutant.   \\ \hline
     $\gamma$ & the amplitude of the fitness fluctuations.  \\ \hline
     $\alpha \equiv s_0/g$ & the ratio between the constant selective force and  the strength of temporal environmental stochasticity  \\ \hline
     $G \equiv N \delta \gamma^2/2$ & scaled environmental stochasticity. \\ \hline
    $T_A(n=1)$ & mean persistence time for a two species system, if at $t=0$ there is only a single mutant $(n=1)$. Average (for this and other quantities) is taken over histories and initial conditions. \\ \hline
    $\Pi(n=1)$ & mean (over initial conditions) chance of fixation for a single mutant.  \\ \hline
    $T_f(n=1)$ & mean  time  to fixation  for a single mutant. \\ \hline
    \end{tabular}

\end{center}

\end{table}

 Our operational definition of a "generation" is $N$ duels. To use the formulas presented below with a different definition of a generation time, say, $AN$ duels, one should stick to the definition of $\delta$ as the persistence time of the environment in units of $N$. For example, if the weather changes every 100 duels and the size of the community is $N = 1000$, $\delta = 0.1$ no matter what $A$ is. Doing that, the formulas obtained here may be used as long as  $T_f$ and $T_A$  are divided by $A$.

 As explained, the results presented here are the outcomes of large-$N$ asymptotic analysis. In particular, the asymptotic matching technique used in the appendices assumes that the demographic noise terms $1/N$ is negligible, with respect to $gx(1-x)$, as long as $x$ is not too close to zero or one. Accordingly, our analysis does not cover the limit in which the environmental stochasticity vanishes, i.e., $g=0$: the value of $g$ may be vanishingly small as long as
 \begin{equation} \label{cond}
 G = Ng \gg 1.
  \end{equation}
   This implies that we cannot recover the purely demographic limit where both $g$ and $s_0$ vanishes. However for any finite $s_0$ our expressions converge to the correct answer even in the limit $g \to 0$ as long as $G \gg 1$  (see discussion below).

   In the following section we compare our results with numerical solutions of the backward Kolomogorov equations that involve simple inversion of the $2N \times 2N$ transition matrix, as detailed in the first appendix of \cite{danino2016stability}. Using the sparsity of the relevant matrices we were able to reach system sizes up to $N = 10^6$.

\section{Results}

 \subsection{The chance of fixation $\Pi(n=1)$}

 For pure demographic noise (neutral system, $s_0 =0$ and $g=0$) the chance of a mutant to win is known to be $$\Pi_{g=s_0=0}(n=1)= \frac{1}{N}. $$ This result is trivial: since all individuals are symmetric, the chance of the lineage of each of them to reach fixation must be equal.

  Under constant selection $s_0$ (still $g=0$) the chance of a single advantageous mutant to reach fixation is
  \begin{equation} \label{ps}
   \Pi_{g=0}(n=1)=\frac{1-e^{-s_0}}{1-e^{-N s_0}} \approx 1-e^{-s_0} \approx s_0,
  \end{equation}
  where the  first approximation is  the strong selection ($Ns_0 \gg 1$) limit,  and the second corresponds to the large $N$, small $s_0$,  limit.

  The intuitive argument behind
 this result is as follows~\cite{desai2007speed}: the mutant lineage starts to feel the deterministic bias only at $n_c(s_0,g=0) \sim 1/s_0$, where its abundance grows on average by one individual per generation. Below $n_{c}$ the process is dominated by the demographic noise. Therefore, the chance of fixation is actually the chance of the lineage of  a single mutant to reach $n_c$ under pure demographic noise  $$\Pi_{g=0}(n=1) \approx \frac{1}{n_c(s_0,g=0)} \approx s_0.$$ The condition for strong selection is translated to $N \gg n_c$.

\begin{figure}
\includegraphics[width=10cm]{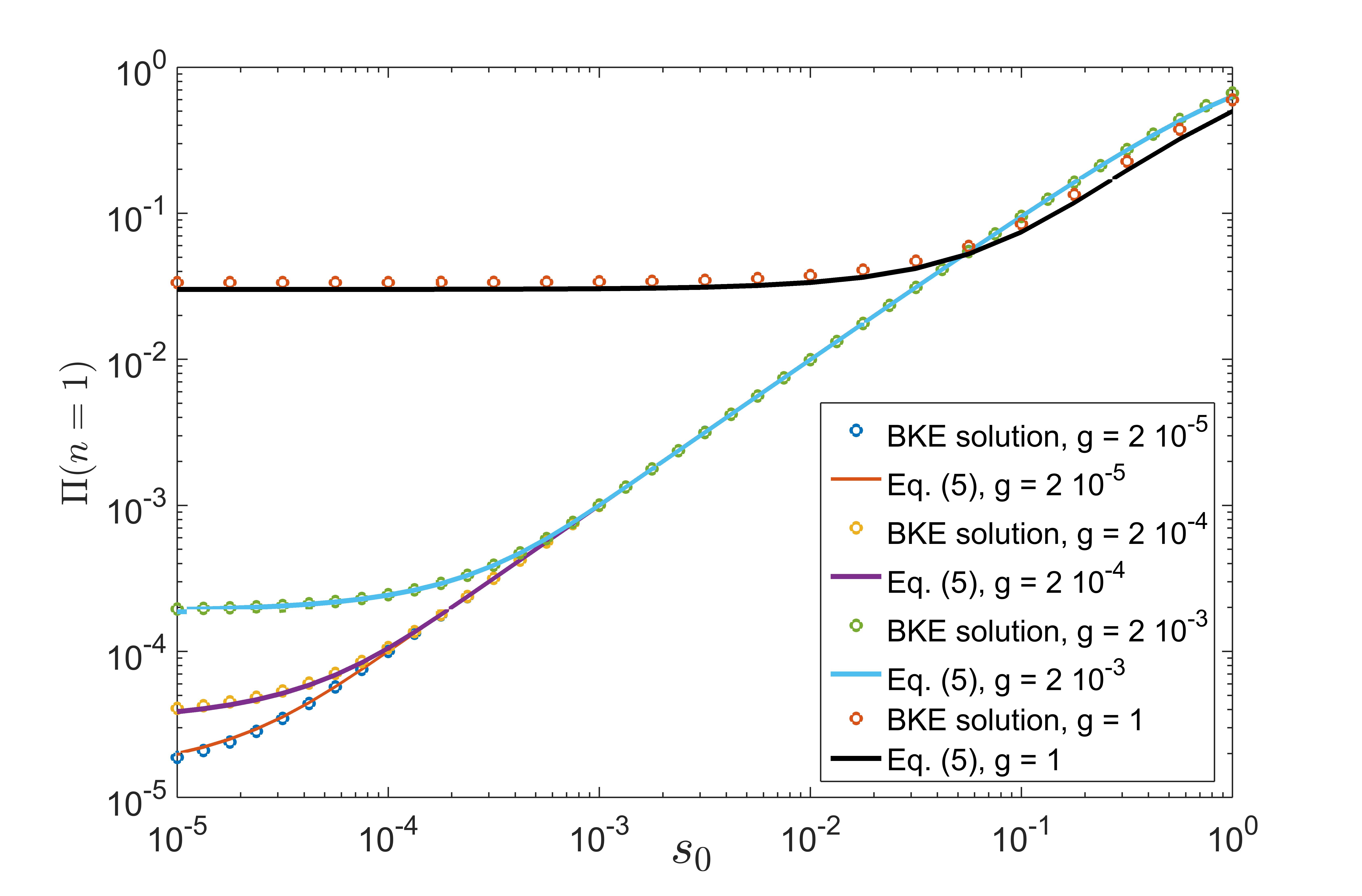}
\caption{\textbf{$\Pi(n=1)$ vs. $s_0$ for different values of $G$.} $\Pi(n=1)$, as obtained from numerical solutions of the discrete backward Kolomogorov equation (open circles), and the large $G$ approximation, Eq. (\ref{eq1}) (full lines), are plotted against $s_0$ for $N = 10^5$ and different values of $G= N \gamma^2 \delta/2$ (see legend). The fit is almost perfect, with only slight deviations (where the analytic formula still have the same shape) at $g=2 \cdot 10^{-5}, \ G=2$  (see Eq. \ref{cond})  and at $g = 1, \ G = 10^5$ (where the continuum approximation becomes problematic). For small values of $s_0$, the chance of fixation grows as $g$ increases. On the contrary, for large values of $s_0$ the chance of fixation decreases when $g$ increases, as explained in the text.} \label{fig1}
\end{figure}

 Now let us turn to our results. For a single mutant where $g$ is finite and $G = gN \gg 1$, the chance of fixation (calculated in Appendix \ref{apb}) is, \cite{danino2017environmental},
 \begin{equation} \label{eq1}
\Pi(n=1) \sim \frac{1-\frac{1}{(1+g)^{s_0/g}}}{1-G^{-2s_0/g}}.
\end{equation}
As demonstrated in Figure \ref{fig1}, this formula matches almost perfectly, without any fitting parameters, the numerical solutions of the discrete, exact BKE. Slight deviations are observed at $G=2$, where the asymptotic matching analysis becomes problematic [see  Eq. (\ref{cond})]. When the noise is very large ($g=1$) tiny deviations are observed again, here the reason is that the continuum approximation fails close to $x=0$ and $x=1$ (see discussion section).

The formula  for $\Pi(n=1)$, given in Eq. (\ref{eq1}) has the following features:

\begin{itemize}

  \item For $s_0 = 0$, $g$ finite,  Eq. (\ref{eq1}) converges to the expression suggested in \cite{cvijovic2015fate}, namely,
      \begin{equation} \label{eq80}
 \Pi(n=1)= \frac{\ln(1+g)}{2\ln(Ng)}.
\end{equation}
 In this case the chance  of fixation \emph{increases} with $g$. To understand why, note that in the large $N$ limit under environmental stochasticity the abundance preforms a random walk along the logarithmic-abundance axis, so the chance of fixation is much larger than  $1/N$ (the chance in the purely demographic case) since the mutant lineage may conquer the whole system in ${\cal O}(log N)$ steps.

 \item On the other hand, when $s_0$ and $g$ are finite but $N \to \infty$, the denominator in Eq. (\ref{eq1}) is unity and $\Pi(n=1)$ is a monotonously \emph{decreasing} function of $g$.  This has to do with the value of $n_{c}$, below which the system is dominated by noise and above it the growth is deterministic and fixation occurs almost surely. While for a system without environmental variations $n_c(s_0,g=0) = 1/s_0$, when $g>0$ \cite{cvijovic2015fate,danino2017environmental}
\begin{equation} \label{eq5}
n_{c} (s_0,g)  = \frac{e^{g/s_0}-1}{g}.
\end{equation}
This expression converges to $1/s_0$ when $g=0$, but increases exponentially with $g$ so it is more difficult for the mutant lineage to enter the deterministic growth zone.

\item Accordingly, for any finite value of $N$ and $s_0$ there is a critical strength of environmental stochasticity, $g_c$,  above which the chance of fixation increases with $g$. At $g_c$   $d\Pi_{n=1}/dg$  vanishes: this yields a transcendental equation for the critical noise level. While we cannot solve for $g_c$ in general, numerical solutions seem to indicate that $g_c \approx s_0 \ln(N)$. Up to logarithmic corrections one may obtain this expression from the condition $N = n_{c} (s_0,g)$, so the system is close to its time-averaged neutral limit (in the sense used in \cite{kalyuzhny2015neutral})  when $N < n_c$ and is in the  (strong selection) regime for $N>n_c$.
    
    This outcome may have interesting implications to the theory of bet-hedging strategies, phenotypic plasticity and related phenomena \cite{kussell2005phenotypic,philippi1989hedging}. Bet-hedging allows species and individuals to cope with changing environmental conditions  by decreasing their fitness in their typical conditions in exchange to increased fitness under stressful conditions. If such a strategy happens to increase the time average log fitness $s_0$ then of course it reduces the chance of extinction. However, if the only effect of these strategies is to reduce the variance in fitness $\gamma$ while keeping $s_0$ fixed, they will be beneficial for a species in a zero-sum competitive community only in the strong selection limit.

    \item For deleterious mutations ($s_0<0$) the chance of fixation decays with $N$ like a power-law, $(Ng)^{-2|s_0|/g}$.

\item As discussed towards the end of the methods section, the case $g=0$ is problematic since the condition $G \gg 1$ no longer holds. Still, as long as $s_0$ is  finite, taking the limit $g \to 0$ is legitimate if $G$ is still large, e.g., $g \sim 1/\sqrt{N},$ as $ \ N \to \infty$. In this case Eq. (\ref{eq1}) converges to the large  $N$ limit of a system with constant selection, $1-e^{-s_0} \approx s_0$, as needed.

    \item As explained above, the pure demographic noise result $\Pi = 1/N$ cannot emerge from Eq. (\ref{eq1}) by taking both $s_0$ and $g$ to zero. Since $N$ should be taken to infinity first, the chance of fixation in this case vanishes. Under constant selection the chance of fixation is finite even in the infinite $N$ limit (this is why we obtained the correct result in that case), but not under pure demographic stochasticity. However, in almost any realistic scenario either $s_0$ or $g$ (and perhaps both) are larger than $1/N$. If both $s_0$ and $g$ are vanishingly small  one may simply use the results for the pure demographic scenario since the selective forces are only tiny perturbation.
\end{itemize}

 \subsection{The time to absorption $T_A$}

\begin{figure}
\includegraphics[width=14cm]{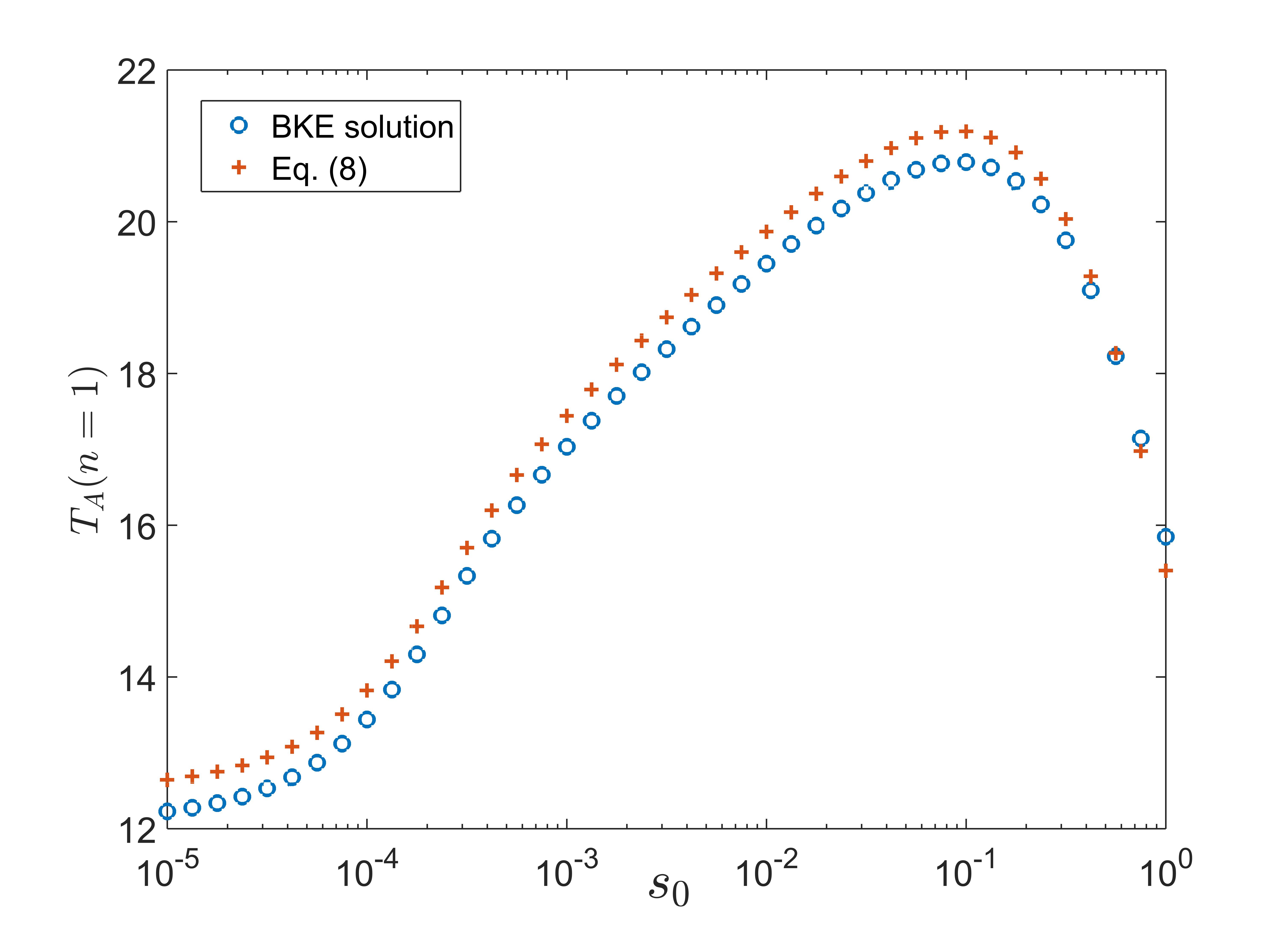}
\caption{$T_A(n=1)$ vs. $s_0$. The time to absorption of a single mutant, as obtained from numerical solutions of the discrete backward Kolomogorov equation (open circles), is compared with the predictions of Eq. (\ref{eq2z}) (plus signs). The results  are plotted against $s_0$ for $N = 10^5$, $\gamma = 0.1$ and $\delta = 0.09$ ($g = 4.5 \cdot 10^{-4}$). Since the extinction times  are order one, the large $N$ behavior of  $T_A $ is determined by $\Pi \cdot T_f$. Accordingly $T_A$ first increases with $s_0$ (since $\Pi$ increases) and then decreases (when the dominant effect is the decrease of $T_f$ with $s_0$).    }\label{fig2}
\end{figure}

 The time to absorption $T_A$ is the average time between the event of mutation/invasion until the system becomes homogenous again, i.e., until the mutant lineage either goes extinct or reaches fixation. In Appendix \ref{apc} we show  that the asymptotic expression for this quantity is,

 \begin{equation} \label{eq2z}
T_A(n=1) =\left(\frac{\ln(G)}{s_0}-\frac{G^{2s_0/g}\beta_2-\beta_1}{G^{2s_0/g}-1}\right) \left( 1-\frac{1}{(1+g)^{s_0/g}} \right)-\frac{1}{g(1+g)^{s_0/g}} \int_0^{g} \frac{\ln(z) \ dz}{(1+z)^{1-s_0/g}},
\end{equation}
where  $$ \beta_1 \equiv \frac{1}{s_0} [H(\alpha)+\pi ctg(\pi \alpha)+\ln(G)],$$
 $$ \beta_2 \equiv \frac{1}{s_0} [-H(-\alpha)+\pi ctg(\pi \alpha)-\ln(G)],$$ and $H(x)$ is the Harmonic number.  The expression in Eq. (\ref{eq2z}) is quite complicated but becomes very simple as $N$ approaches infinity, where it takes the form,

\begin{equation} \label{eq2}
 T_A(n=1)  \sim \frac{2}{s_0}\left( 1-\frac{1}{(1+g)^{s_0/g}}\right) \ln(N).
 \end{equation}
 However, the rate of convergence of Eq. (\ref{eq2z}) to Eq. (\ref{eq2}) is slow, and when we tested our results against the numerical solutions of the BKE at $N=10^5$ (Figure \ref{fig2}), we implemented Eq. (\ref{eq2z}).

\begin{itemize}
\item When $g \to 0$ Eq. (\ref{eq2z}) converges to
\begin{equation} \label{eq2a}
T_A(n=1)  \sim 2 (1-e^{-s_0}) \ln(N)/s_0 \approx 2 \ln(N).
 \end{equation}
 This is the correct limit for a singleton without environmental stochasticity \cite{ewens2012mathematical}.

\item On the other hand when $s_0 \to 0$ (\ref{eq2}) yields
 \begin{equation} \label{eq2b}
T_A(n=1)  \sim  2 \  \frac{\ln(1+g)}{g} \ln(N),
  \end{equation}
  which it the result obtained in \cite{danino2016stability}. The simple expression  (\ref{eq2}) interpolates  between these two limits. Unlike the fixation time (see below),  $T_A$ is always logarithmic in $N$, hence the interpolation between these two limits involves only the prefactor.

\end{itemize}

 \subsection{The time to fixation $T_f$}

\begin{figure}
\includegraphics[width=12cm]{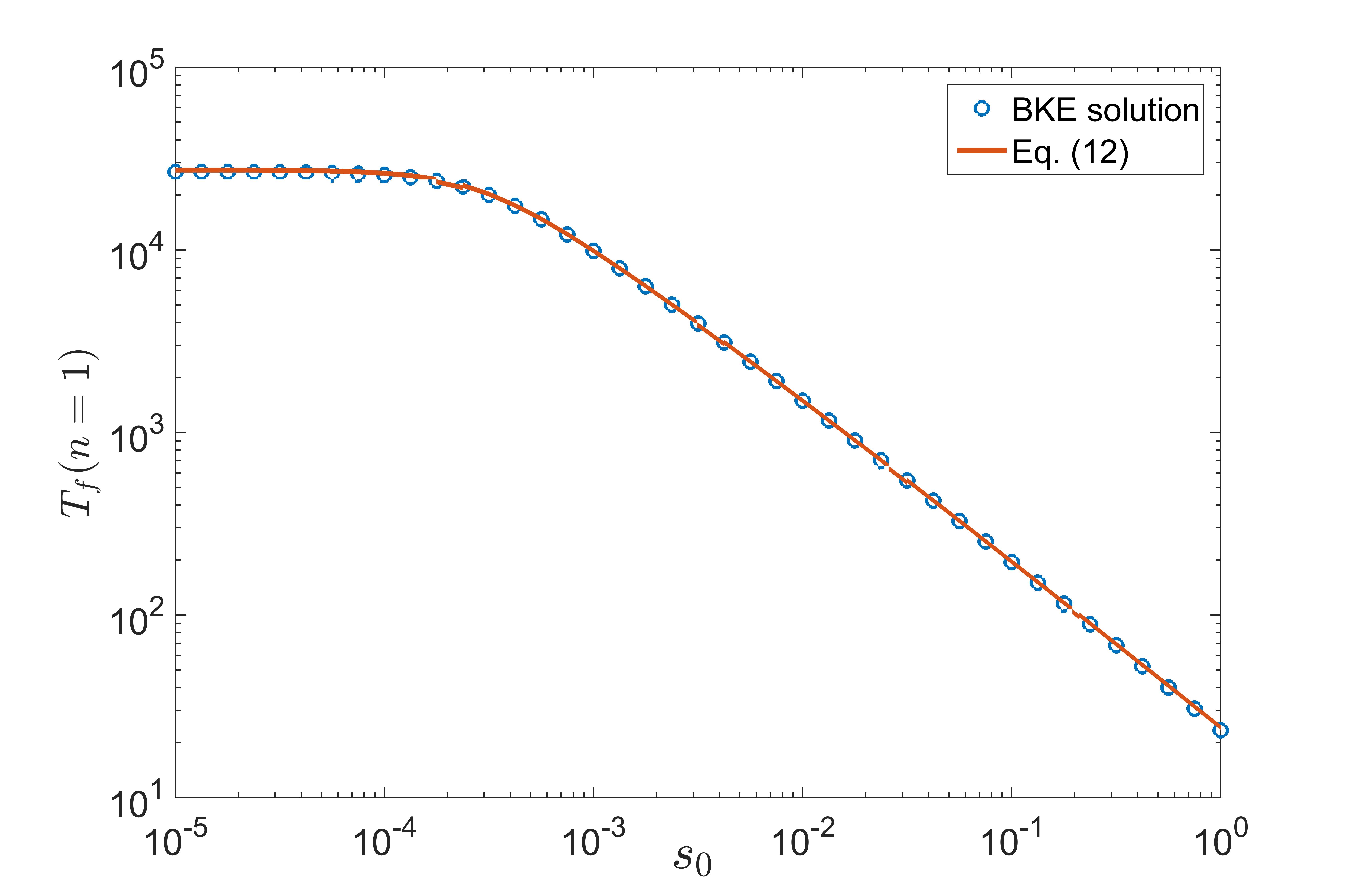}
\caption{$T_f(n=1)$ vs. $s_0$. The time to fixation, as obtained from numerical solutions of the discrete backward Kolomogorov equation (open blue circles), is compared with the predictions of Eq. (\ref{eq3}) (full red line), both plotted against $s_0$ for $N = 10^5$, $\gamma = 0.1$ and $\delta = 0.09$ ($g = 4.5 \cdot 10^{-4}$).}\label{fig3}
\end{figure}

 The fixation time is the average time between mutation and fixation, when the average is taken over all the trajectories that start at $n=1$ and end up at $n=N$. In Appendix \ref{apd} we show that,
 \begin{equation}\label{eq3}
T_f (n=1) \sim 2 \left( \frac{[1+G^{2s_0/g}]\ln(G)}{s_0[G^{2s_0/g}-1]}-\frac{\pi ctg(\pi s_0/g)}{s_0} + \frac{H(s_0/g)+G^{2s_0/g}H(-s_0/g)}{s_0[G^{2s_0/g}-1]} \right).
\end{equation}
Although this expression has a singular point at $s_0=g$, the curve is smooth  out of  a region of width  $1/N$ around the singular point, so this singularity is negligible in the  large $N$ limit. Figure \ref{fig3} depicts the fit of (\ref{eq3}) to the numerical solution of the discrete BKE.

\begin{itemize}
\item As $N \to \infty$ when  $s_0$ and $g$ are kept fixed, Eq.  (\ref{eq3}) yields,
\begin{equation} \label{eq209}
T_f (n=1) \sim \frac{2}{|s_0|} \ln(N),
\end{equation}
as expected. In this limit the random walk in the log-abundance space, associated with the environmental stochasticity, may be neglected with respect to the constant bias.

\item On the other hand, for fixed $N$ and $g$ when $s_0$ vanishes,
\begin{equation} \label{eq210}
T_f (n=1) \sim \frac{2}{3g} \ln^2(gN).
\end{equation}
Note that  Eq. (\ref{eq210}) is a result of a third order expansion of (\ref{eq3}), where the lower order terms exactly cancel each other. For  fixed $g$, the large $N$   approximation [Eq.  (\ref{eq209})] is valid (as in the case of $\Pi$) as long as $G^{2s_0/g} \gg 1$, i.e., as long as $s_0 \ln(gN)/g > 1$, or simply $N \gg n_c(s_0,g)$.

 \item $T_f$ is a symmetric function of $s_0$ (this was shown, for a model without environmental stochasticity, in \cite{taylor2006symmetry}. Eq. (\ref{eq3}) implies that this feature holds under fluctuating selection). $T_f$ peaks at $s_0 = 0$.
 \end{itemize}

\section{Discussion}

Through this paper we have calculated and analyzed three  fundamental quantities that have to do with the fate of a  mutant, or an immigrant, in a community of size $N$ under the effect of  selection, demographic stochasticity (drift) and environmental variations.  These quantities: the chance of fixation, the time to absorption and the time to fixation, govern the dynamics of evolution for a community with fixed mutation rate, as explained in \cite{cvijovic2015fate,danino2017environmental}. We have focused our discussion on the fate of a single  mutant/invader; other quantities, like the \emph{maximum} time to absorption (for example, in the absence of selection it is clear that the maximum time to absorption occurs when the community is divided equally between the two species, $n(t=0) = N/2$) were calculated in  \cite{danino2016stability}. In fact, the same analytic methods we have used in the appendices may be utilized to calculate the relevant quantities for any value of $n$ (not only a single mutant) given $N$, $g$ and $s_0$.

Our analysis is based on equations for the average quantities, where average is taken over both histories and initial conditions. These equations are similar to those obtained using the standard diffusion approximation \cite{karlin1981second}, but there are a few technical differences. Our treatment begins with the introduction of an  exact backward Kolomogorov equation, followed by transition to the continuum, dominant balance analysis that allows us to neglect a few terms and then by the calculation of the large $N$ asymptotic behavior. This methodology allows for better identification of the limits of our theory. We would like to emphasize three of these limiting factors:

\begin{enumerate}
  \item \textbf{"Single sweep" fixation}: In our results, the effect of environmental noise is expressed by a single parameter $g=\gamma^2 \delta/2$. This parameter may be considered as the diffusion constant in the log-abundance space: If $x=n/N$ is the fraction of mutants and $\dot{x} = \pm \gamma x(1-x)$, the system performs an unbiased random walk on the  $z=\ln[x/(1-x)]$ axis with an effective ``diffusion constant" $\gamma^2 \delta$.   Clearly, this is not the case when the takeover of the community takes place during $\delta$ generations, i.e., when $\delta > \ln(N)/(\gamma+s)$. In such a case the single parameter ($g$) scaling breaks down.   This possibility has been discussed in \cite{cvijovic2015fate,danino2016stability}, but appears to be less interesting as it describes an isolated catastrophe instead of the accumulation  of environmental variations over time.

  \item \textbf{Breakdown of the continuum approximation}: When the quantities considered here change their values abruptly between $n$ and $n+1$ (this happens, usually, close to $n=0$ or $n=N$) the transition to the continuum may fail and one should consider the original difference equations instead of the differential equations. For a detailed discussion of this problem (in different system) and a WKB recipe suggested for that case, see \cite{kessler2007extinction}. In Fig. \ref{fig1} above this problem manifests itself in the $g=1, \ G = 2 \cdot 10^5$ (very strong stochasticity) case.

  \item \textbf{Breakdown of the asymptotic matching}: As discussed above, our asymptotic matching analysis is based on the assumption that $G \gg 1$. If this is not the case, one cannot identify the inner, middle and outer regimes as done in the appendices. In Fig. \ref{fig1} we have seen, indeed, that when $G=2$ the deviations of our theory from the exact numerical results are identifiable.
\end{enumerate}

One aspect of community dynamics that we did not take into account is the stabilizing mechanism known  as the  storage effect \cite{chesson1981environmental}. For a system with storage, the environmental variations stabilizes the coexistence state (in the absence of selection, at $n=N/2$), thus facilitating the invasion of new species or a mutant (and increasing the chance of fixation \cite{danino2017environmental}). However, quantities like the time to fixation, or even the chance of fixation per se, are less relevant for systems with storage effect. In these systems, when a mutant invades it typically reaches the coexistence  state and stay around it a long time (about $N^{1/\delta}$ generations, see \cite{danino2016stability}), only then one of the species goes extinct. Accordingly, for most purposes the relevant quantity under storage is not the chance of fixation but the chance of establishment~\cite{danino2017environmental}. We hope to address this question in subsequent publication.

\section{\bf Acknowledgments} We acknowledge the support of the Israel
Science Foundation, grant no. $1427/15$.

\bibliography{refs}
\newpage
\appendix

\section{Technical definitions and the Backward Kolomogorov equation} \label{apa}

Our Moran process under dichotomous stochasticity is fully characterized by twelve transition rates. In each elementary step the mutant population may stay the same or grow/shrink by one individual. At the same time the environment may switch from its $(+)$ state (where the chance of the mutant to win a duel is $1/2 + s_0/4 + \gamma/4$) to a $(-)$ state (where the winning probability is $1/2 + s_0/4 - \gamma/4$) and vice versa. Defining $x=n/N$ as the mutant fraction in the population, the transition probabilities $W$ are given by,

\begin{eqnarray}
W^{++}_{n \to n+1} = W^{--}_{n \to n-1}  &=& 2x(1-x)\left(\frac{1}{2}+\frac{s_0}{4}+\frac{\gamma}{4}\right) \left(1-\frac{1}{\delta N}\right)  \nonumber \\ W^{--}_{n \to n+1} = W^{++}_{n \to n-1} &=& 2x(1-x)\left(\frac{1}{2}+\frac{s_0}{4}-\frac{\gamma}{4}\right) \left(1-\frac{1}{\delta N}\right) \nonumber \\
W^{+-}_{n \to n+1}=W^{-+}_{n \to n-1} &=& 2x(1-x)\left(\frac{1}{2}+\frac{s_0}{4}-\frac{\gamma}{4}\right) \frac{1}{\delta N}   \\ W^{-+}_{n \to n+1} = W^{+-}_{n \to n-1}&=& 2x(1-x)\left(\frac{1}{2}+\frac{s_0}{4}+\frac{\gamma}{4}\right)\frac{1}{\delta N} \nonumber \\
W^{++}_{n \to n} = W^{--}_{n \to n}&=& \left(1-\frac{1}{\delta N}\right) [1-2x(1-x)] \nonumber \\ W^{+-}_{n \to n} = W^{-+}_{n \to n}&=& \frac{1}{\delta N}[ 1-2x(1-x)] \nonumber
\end{eqnarray}
where $W^{++}_{n \to n+1}$ is the probability to increase the mutant population by one individual while staying in the plus environment, while $W^{+-}_{n \to n+1}$ is the chance that the environment switches from plus to minus and after this switch the mutant population grew.

After each duel time is incremented by $1/N$, so the BKE for the time to absorption, say, takes the form,
\begin{eqnarray} \label{eq60a}
T_A^+(n) = \frac{1}{N} &+& W^{++}_{n \to n+1}  T_A^+(n+1) + W^{++}_{n \to n-1} T_A^+(n-1) + W^{++}_{n \to n}T_A^+(n) \nonumber \\ &+& W^{+-}_{n \to n-1} T_A^-(n-1) + W^{+-}_{n \to n+1} T_A^-(n+1)+ W^{+-}_{n \to n}T_A^-(n)  \\
T_A^-(n)= \frac{1}{N} &+& W^{--}_{n \to n-1}  T_A^-(n-1) + W^{--}_{n \to n+1} T_A^-(n+1)+W^{--}_{n \to n}T_A^-(n) \nonumber \\ &+& W^{-+}_{n \to n+1}T_A^+(n+1) + W^{-+}_{n \to n-1} T_A^+(n-1)+W^{-+}_{n \to n}T_A^+(n) \nonumber
\end{eqnarray}

Defining $T_A(n) = [T_A^+(n) + T_A^-(n)]/2$, $\Delta(n) = [T_A^+(n) - T_A^-(n)]/2$, moving to the continuum limit and expanding $T(x \pm 1/N)$ to the second order in a  Taylor series  one finds:
\begin{eqnarray} \label{eq61a}
\frac{2 \Delta }{\delta N x(1-x)} &=& \left( 1-\frac{2}{\delta N} \right)\left[ \frac{\gamma}{N} T_A' + \frac{\Delta''}{N^2} + s_0 \frac{\Delta'}{N} \right] \\ \nonumber
-\frac{1}{x(1-x)} &=&  \frac{T_A''}{N}  +\gamma \Delta' + s_0 T_A'.
\end{eqnarray}
Where primes indicate a derivative with respect to $x$. If $\delta$ is kept fixed (say, $1/10$ of a generation) and $N$ increases, $\delta N \gg 1 $ and
 \begin{eqnarray} \label{eq62}
\frac{2 \Delta }{\delta  x(1-x)} &=&  \gamma T_A' + \frac{\Delta ''}{N} + s_0 \Delta' \\ \nonumber
-\frac{1}{x(1-x)} &=&  \frac{T_A''}{N}  +\gamma \Delta' + s_0 T_A'.
\end{eqnarray}
Neglecting the $\Delta''/N$ and the $\Delta'$ terms in the upper equation, solving for $\Delta$ in terms of $T_A'$, $\Delta = \gamma \delta x(1-x) T_A'/2$, and plugging the expression for $\Delta'$ into the lower equation, one obtains,
\begin{equation} \label{eq63}
-\frac{1}{x(1-x)} = [\frac{1}{N}+gx(1-x)]T_A'' + [s_0  + g (1-2x)]T_A'.
\end{equation}
which is exactly Eq. \ref{eq90}.

The dominant balance argument that leads to the neglect  of  the $\Delta''/N$ and the $\Delta'$ terms in the upper equation was motivated by a term by term analysis of the numerical solutions of the discrete BKE (\ref{eq60a}). The argument is self consistent in the middle regime: extracting $\Delta$ from Eq. (\ref{eq102}) and calculating the relevant terms, the two neglected terms were found to be subdominant in the $G \to \infty$ limit.

Apparently, this has to be the case.  First, if the $\gamma S'$ term is subdominant, then $\Delta =0$ and the effect of environmental stochasticity vanishes. Therefore, the only question is, which term balances the $\gamma S'$ in the large $N$ limit. Clearly, if the balancing term is the $\Delta'$  environmental stochasticity only renormalizes the value of $s_0$, while the $\Delta''$ leads to a renormalization of the strength of the demographic noise. Accordingly, and in agreement with the outcomes of our numerical solutions, the dominant balance argument makes sense.

\section{Large-N asymptotics for the chance of fixation $\Pi$}\label{apb}

Defining $x = n/N$ and using the results of  \cite{danino2016stability} (Appendix C), the chance of fixation $\Pi(x)$  satisfies \cite{redner2001guide},
\begin{equation} \label{eq63}
\left(\frac{1}{N}+gx(1-x)\right)\Pi''(x) + (s_0 + g (1-2x)) \Pi'(x)=0, \qquad \Pi(0) = 0 \quad \Pi(1) = 1.
\end{equation}

To calculate the large $N$ asymptotic of $\Pi$, we will solve (\ref{eq63}) in three different regions:

\begin{enumerate}
  \item The inner region $0 \leq x<<1$. In this region the number of individuals may be small [even for large $N$, $n=Nx$ may be ${\cal O}(1)$] and demographic noise affects the system. The relevant equation for $\Pi_{in}(x)$ is obtained from (\ref{eq63}) by replacing $1-x$ and $1-2x$ by unity, and is subject to a single boundary condition at zero,

      \begin{equation} \label{eq64}
\left(\frac{1}{N}+gx\right)\Pi_{in}''(x) + (s_0 + g) \Pi_{in}'(x)=0, \qquad \Pi_{in}(0) = 0.
\end{equation}
Using an integrating factor one may  easily show that,
 \begin{equation} \label{eq65}
\Pi_{in}(x) = C_1 \left( 1-\frac{1}{(1+Gx)^\alpha} \right),
\end{equation}
where $\alpha \equiv s_0/g$ and $G \equiv Ng$. Eq. (\ref{eq65}) satisfies the left boundary condition and depends on one constant, $C_1$, to be determined below using an asymptotic matching.

\item In the intermediate region, $0 \ll x \ll 1$, the demographic noise is negligible  (for any  $x$ in this regime,  when $N \to \infty$ the $1/N$ term is much smaller than  $gx(1-x)$. Accordingly, the relevant equation is,
    \begin{equation} \label{eq66}
\left(gx(1-x)\right)\Pi_{M}''(x) + (s_0 + g(1-x)) \Pi_{M}'(x)=0,
\end{equation}
or,
 \begin{equation} \label{eq67}
\Pi_{M}''(x) + \left( \frac{s_0}{g} \ln'\left(\frac{x}{1-x}\right) + \ln'(x[1-x])\right) \Pi_{M}'(x)=\left(\Pi_{M}' \left(\frac{x^{\alpha+1}}{(1-x)^{\alpha-1}}\right)\right)'= 0.
\end{equation}
This yields,
 \begin{equation} \label{eq68}
\Pi_{M}(x) = C_2 \left( \frac{1-x}{x} \right)^\alpha +C_3.
\end{equation}
Here we have two free constants as none of the boundary condition is relevant in the middle regime.

\item Finally, in the outer regime $1-x \ll 1$, $x$ is close to one and $1-2x \approx (-1)$,  so we have to consider
\begin{equation} \label{eq69}
\left(\frac{1}{N}+g(1-x)\right)\Pi_{out}''(x) + (s_0 - g) \Pi_{out}'(x)=0, \qquad \Pi_{out}(1) = 1.
\end{equation}
obtaining,
\begin{equation} \label{eq70}
\Pi_{out}(x) = 1 - C_4 \left(1- [1+G(1-x)]^\alpha \right).
\end{equation}
In fact, the expression (\ref{eq70}) may be obtained directly from (\ref{eq65}) using the symmetry of the problem: the chace of a species of abundance $x$ and selection parameter $s_0$ to win, is the same as its chance to lose  if its abundance is $1-x$ and the selective parameter is reversed,
\begin{equation} \label{eq70}
\Pi(s_0,x) = 1 - \Pi(-s_0, 1-x).
\end{equation}

\end{enumerate}

Using equations (\ref{eq65}),(\ref{eq68}) and (\ref{eq70}), we can now find the $C$ constants by matching the solutions in the overlap regimes. $Pi_{in}$ must match $\Pi_M$ when $x \ll 1$ but $Gx \gg 1$, meaning that
\begin{equation}
C_1 - \frac{C_1}{(Gx)^\alpha} =\frac{ C_2}{x^\alpha} +C_3 \qquad C_1 = C_3, \ \ G^\alpha C_2 = -C_1.
\end{equation}
A similar matching of $\Pi_M$ and $\Pi_{out}$ when both $1-x \ll 1$ and $G(1-x) \gg 1$ yields
\begin{equation}
1-C_4 + C_4 [G(1-x)]^\alpha = C_2 (1-x)^\alpha +C_3 \qquad 1-C_4 = C_3, \ \ G^\alpha C_4 = C_2.
\end{equation}
Accordingly,
\begin{eqnarray}\label{eq11a}
 C_1 = C_3 &=&  \frac{1}{1-G^{-2\alpha}} \nonumber \\
C_2 &=& \frac{1}{G^{-\alpha}-G^{\alpha}}  \\
 C_4 &=&  \frac{1}{1-G^{2\alpha}} . \nonumber
\end{eqnarray}

In the large $N$ limit,
\begin{eqnarray}\label{eq12}
\Pi_{in}(x) &\sim& \left( \frac{1}{1-G^{-2\alpha}} \right) \left( 1-\frac{1}{(1+Gx)^\alpha} \right), \nonumber \\
\Pi_{M}(x) &\sim& \left( \frac{1}{G^{-\alpha}-G^{\alpha}}  \right) \left( \frac{1-x}{x} \right)^\alpha +\frac{1}{1-G^{-2\alpha}},  \\
\Pi_{out}(x) &\sim& 1 - \left( \frac{1}{1-G^{2\alpha}}  \right) \left(1- [1+G(1-x)]^\alpha \right). \nonumber
\end{eqnarray}
The chance of a single mutant ($n=1$, $x=1/N$, $Gx =Ngx = g$) to win  is given by,
\begin{equation}
\Pi_{in}(1/N) = \Pi(n=1) \sim \frac{1-\frac{1}{(1+g)^{s_0/g}}}{1-(Ng)^{-2s_0/g}} .
\end{equation}

\section{Absorption times} \label{apc}

The relevant BKE is,
 \begin{equation} \label{eq90}
\left(\frac{1}{N}+gx(1-x)\right)T_A''(x) + (s_0 + g(1-2x)) T_A'(x)= -\frac{1}{x(1-x)},  \qquad T_A(0) = T_A(1)  = 0.
\end{equation}

In the inner regime $x \ll 1$
\begin{equation} \label{eq91}
\left(\frac{1}{N}+gx\right)T_{A,in}''(x) + (s_0 + g) T_{A,in}'(x)= -\frac{1}{x},  \qquad T_{A,in}(0)   = 0.
\end{equation}
Accordingly
\begin{equation} \label{eq92}
\left(T_{A,in}'\left(\frac{1}{N}+gx\right)^{1+\alpha} \right)' =\frac{\left(\frac{1}{N}+gx\right)^{\alpha}}{x}.
\end{equation}
The solution that satisfies the left boundary condition is,
\begin{equation} \label{eq93}
T_{A,in}(x) = \tilde{C}_1 \left( 1-\frac{1}{(1+Gx)^\alpha} \right)-N \int_0^x \frac{dt}{\left(1+Gt\right)^{1+\alpha}} \int^t \ dq \frac{\left(1+Gq\right)^{\alpha}}{q}.
\end{equation}
The inner integral may be written as
\begin{equation} \label{eq94}
\int^t \ dq (1+Gq)^{\alpha}\frac{d}{dq}\ln(q) = (1+Gt)^\alpha \ln(t) - \alpha G \int^t (1+Gq)^{\alpha-1} \ln(q) dq
\end{equation}
Plugging (\ref{eq94}) into (\ref{eq93}) and using integration by parts to simplify, one obtains,
\begin{equation} \label{eq95}
- \alpha G \int_0^x \frac{dt}{(1+Gt)^{1+\alpha}} \int^t (1+Gq)^{\alpha-1} \ln(q) dq = \frac{1}{(1+Gx)^{\alpha}} \int_0^x \frac{\ln(t)}{(1+Gt)^{1-\alpha}} \ dt - \int_0^x \frac{\ln(t)}{(1+Gt)}.
\end{equation}
The last term of (\ref{eq95}) cancels with the middle term of  (\ref{eq94}) when they both plugged in (\ref{eq93}). Accordingly,
\begin{equation} \label{eq96}
T_{A,in}(x) =\tilde{C}_1 \left( 1-\frac{1}{(1+Gx)^\alpha} \right)-\frac{N}{(1+Gx)^{\alpha}} \int_0^x \frac{\ln(t)}{(1+Gt)^{1-\alpha}} \ dt.
\end{equation}
 The substitution $z=Gt$ yields,
\begin{equation} \label{eq97}
 \int_0^x \frac{\ln(t)}{(1+Gt)^{1-\alpha}} \ dt = \int_0^{Gx} \frac{\ln(z)-\ln(G)}{(1+z)^{1-\alpha}} \ \frac{dz}{G} = -\frac{\ln(G)}{\alpha G} \left((1+Gx)^\alpha -1\right)+\frac{1}{G}\int_0^{Gx} \frac{\ln(z) \ dz}{(1+z)^{1-\alpha}},
\end{equation}
so finally,
\begin{equation} \label{eq98}
T_{A,in}(x) =\left(\tilde{C}_1+\frac{\ln(G)}{s_0}\right) \left( 1-\frac{1}{(1+Gx)^\alpha} \right)-\frac{1}{g(1+Gx)^{\alpha}} \int_0^{Gx} \frac{\ln(z) \ dz}{(1+z)^{1-\alpha}}.
\end{equation}

To match $T_{A,in}$ with $T_{A,M}$ one needs its asymptotic behaviour as $Gx \to \infty$. Expanding (\ref{eq98}) one finds,
\begin{equation} \label{eq99}
T_{A,in}(Gx \to \infty) \sim \tilde{C}_1 +\frac{g}{s_0^2} -\frac{\ln(x)}{s_0} -\frac{1}{s_0}\frac{H(\alpha)+\pi ctg(\pi \alpha)+\ln(G)}{(Gx)^\alpha} - \frac{\tilde{C}_1}{(Gx)^\alpha}.
\end{equation}
Using the symmetry $T_A(s_0,x) = T_A(-s_0,1-x)$ one can find easily the relevant asymptotic behavior of $T_{out}$,
 \begin{equation} \label{eq100}
T_{A,out}(G(1-x) \to \infty) \sim \tilde{C}_4 +\frac{g}{s_0^2} +\frac{\ln(1-x)}{s_0} +\frac{1}{s_0}\frac{H(-\alpha)-\pi ctg(\pi \alpha)+\ln(G)}{(Gx)^\alpha} - \tilde{C}_4[G(1-x)]^\alpha.
\end{equation}
The expressions (\ref{eq99}) and (\ref{eq100}) should match the intermediate solution $T_{A,M}$ in the relevant regimes. $T_{A,M}$ satisfies,
 \begin{equation} \label{eq101}
gx(1-x)T_{A,M}''(x) + [s_0 + g(1-2x)] T_{A,M}'(x)= -\frac{1}{x(1-x)},
\end{equation}
and admits a relatively simple solution
\begin{equation} \label{eq102}
T_{A,M}(x) = {\tilde C}_3 +{\tilde C}_2 \left(\frac{1-x}{x}\right)^\alpha-\frac{1}{s_0}\ln\left(\frac{x}{1-x}\right).
\end{equation}
Matching in the regime $x \ll 1 \ll Gx$ one finds
\begin{equation} \label{eq102a}
 {\tilde C}_3  = {\tilde C}_1 +\frac{g}{s_0^2}, \qquad {\tilde C}_1+\beta_1 = -  G^\alpha {\tilde C}_2,
 \end{equation}
 where $$ \beta_1 = \frac{1}{s_0} [H(\alpha)+\pi ctg(\pi \alpha)+\ln(G)].$$
 Similarly in the regime $1-x \ll 1 \ll G(1-x)$ the matching yields
\begin{equation}
 {\tilde C}_3  = {\tilde C}_4 +\frac{g}{s_0^2}, \qquad {\tilde C}_4+\beta_2 = -  G^{-\alpha} {\tilde C}_2,
 \end{equation}
 with $$ \beta_2 = \frac{1}{s_0} [-H(-\alpha)+\pi ctg(\pi \alpha)-\ln(G)].$$
 From these algebraic relations one finds,
 \begin{equation} \label{105}
 {\tilde C}_1 =-\frac{G^{2\alpha}\beta_2-\beta_1}{G^{2\alpha}-1}
 \end{equation}

For a single mutant, the time to absorption $T_{A,in}(1/N)$ is obtained by plugging ${\tilde C}_1$ into (\ref{eq98}) with $x \to 1/N$,
\begin{equation} \label{eq106}
T_{A,in}(1/N) =\left(\frac{\ln(G)}{s_0}-\frac{G^{2s_0/g}\beta_2-\beta_1}{G^{2s_0/g}-1}\right) \left( 1-\frac{1}{(1+g)^{s_0/g}} \right)-\frac{1}{g(1+g)^{s_0/g}} \int_0^{g} \frac{\ln(z) \ dz}{(1+z)^{1-s_0/g}}.
\end{equation}
The leading behavior of the time to absorption for a single mutant  is given by the large $N$ asymptotics of (\ref{eq106}),
\begin{equation} \label{eq107}
T_A(n=1) \sim \frac{2}{s_0}\left( 1-\frac{1}{(1+g)^{s_0/g}}\right) \ln(N).
\end{equation}

\section{Time to fixation $T_f$} \label{apd}
To obtain the time to fixation \cite{redner2001guide},  one should solve a BKE  for
\begin{equation} \label{eq200a}
Q(x) = \Pi(x) T_f(x).
\end{equation}

This BKE takes the form,

 \begin{equation} \label{eq200}
\left(\frac{1}{N}+gx(1-x)\right)Q''(x) + (s_0 + g(1-2x)) Q'(x)= -\frac{\Pi(x)}{x(1-x)},  \qquad Q(0) = Q(1)  = 0.
\end{equation}
We would like to solve for $Q$ in the inner, outer and intermediate regime, using the values of $\Pi$ obtained in Eqs. (\ref{eq11a} - \ref{eq12}) for each of these regimes.
 \begin{eqnarray} \label{eq202}
\left(\frac{1}{N}+gx\right)Q_{in}''(x) + (s_0 + g) Q_{in}'(x)&=& -\frac{C_1}{x} + \frac{C_1}{N^\alpha x \left(\frac{1}{N}+gx\right)^\alpha}  \qquad Q_{in}(0) = 0 \nonumber \\ gx(1-x)Q_{M}''(x) + [s_0 + g(1-2x)] Q_{M}'(x)&=& -\frac{C_3}{x(1-x)} - \frac{C_2(1-x)^{\alpha-1}}{ x^{\alpha+1}} \\ \left(\frac{1}{N}+g(1-x)\right)Q_{out}''(x) + (s_0 -g) Q_{out}'(x)&=& -\frac{1-C_4}{1-x} - \frac{C_4 N^\alpha \left(\frac{1}{N}+g(1-x)\right)^\alpha }{ 1-x }  \qquad Q_{out}(1) = 0 \nonumber.
\end{eqnarray}

Since (\ref{eq200}) is linear, the solution for  $Q(x)$ in each regime contains a homogenous term which is equal to $\Pi$ up to a constant, a special solution that has the form of $T$ and another special solution that comes from the last terms if (\ref{eq202}). Denoting the constants of the homogenous solutions by $\overline{C}$, we obtained, for example,
\begin{eqnarray} \label{eq204}
Q_{in}(x) = \overline{C}_1 \left(1-\frac{1}{(1+Gx)^\alpha}\right) - \frac{C_1 N}{(1+Gx)^\alpha} \int_0^x \frac{dt \ \ln(t)}{(1+Gt)^{1-\alpha}} + C_1 N \int_0^x \frac{dt \ \ln(t)}{(1+Gt)^{1+\alpha}} \nonumber \\ Q_{M}(x) =  \overline{C}_2 \left( \frac{1-x}{x} \right)^\alpha +\overline{C}_3 -\frac{C_3}{s_0} \ln\left(\frac{x}{1-x}\right) + \frac{gC_2}{s_0^2} \left( \frac{1-x}{x}\right)^\alpha +\frac{C_2}{s_0} \left( \frac{1-x}{x}\right)^\alpha \ln\left(\frac{x}{1-x}\right) \\ Q_{out}(x) = \overline{C}_4 \left(1-[1+G(1-x)]^\alpha\right) - \frac{1-C_4 N}{(1+G(1-x))^{-\alpha}} \int_0^{1-x} \frac{dt \ \ln(t)}{(1+Gt)^{1+\alpha}} - C_4 N \int_0^{1-x} \frac{dt \ \ln(t)}{(1+Gt)^{1-\alpha}}. \nonumber
\end{eqnarray}
To match these solutions in the overlap regimes, the constant $\overline{C}$ should satisfy,
\begin{eqnarray} \label{eq205}
\overline{C}_3 = \overline{C}_1 +C_1 \frac{g}{s_0^2} + C_1 \beta_2 \qquad \overline{C}_3 = \overline{C}_4 +(1-C_4) \frac{g}{s_0^2} - C_4 \beta_1 \nonumber \\
-G^\alpha \overline{C}_2 = \overline{C}_1  + C_1 \beta_1 \qquad -G^{-\alpha} \overline{C}_2 = \overline{C}_4  + (1-C_4)\beta_2,
\end{eqnarray}
which implies
\begin{equation}
\overline{C}_1 = \frac{2 \beta_1 G^{-2 \alpha} - 2 \beta_2 G^{2 \alpha}}{(G^{2 \alpha}-G^{-2 \alpha})^2}.
\end{equation}
Plugging this into the expression for $Q_{in}$ and evaluating $T_f = Q_{in}/\Pi_{in}$ at $x=1/N$, one finds the fixation time of a singleton:
\begin{equation}\label{eq208}
T_f(1/N) = T_f(n=1) \sim 2 \left( \frac{[1+G^{2 \alpha}]\ln(G)}{s_0[G^{2\alpha}-1]}-\frac{\pi ctg(\pi \alpha)}{s_0} + \frac{H(\alpha)+G^{2\alpha}H(-\alpha)}{s_0[G^{2\alpha}-1]} \right).
\end{equation}

\end{document}